\documentclass{iau}
\usepackage{amsmath}
\usepackage{graphicx}
\usepackage{natbib}

\newcommand{\bootes}{Bo\"{o}tes}
\newcommand{\xbootes}{XBo\"{o}tes} 
\newcommand{\lir}{L_{\rm IR}}
\newcommand{\lx}{L_{\rm x}}

\title[SFR-average BHAR correlation in SF galaxies] 
{A correlation between star formation rate and average black hole
  accretion rate in star forming galaxies}
\author[Chien-Ting J. Chen \& Ryan C. Hickox]
{Chien-Ting J. Chen$^{1,2}$
 \and Ryan C. Hickox$^1$}
\affiliation{$^1$Department of Physics and Astronomy, Dartmouth
  College, 6127 Wilder Laboratory, Hanover, NH 03755, USA\\[\affilskip]
$^2$email: {\tt ctchen@dartmouth.edu}}
\pubyear{2014}
\volume{304}  
\pagerange{xxx--xxx}
\jname{Multiwavelength AGN Surveys and Studies}
\editors{A. Mickaelian, F. Aharonian \& D. Sanders, eds.}
\begin{document}

\maketitle

\begin{abstract}
We present the results of recent studies on the co-evolution of
galaxies and the supermassive black holes (SMBHs) using Herschel
far-infrared and Chandra X-ray observations in the \bootes\ survey
region. For a sample of star-forming (SF) galaxies, we find a strong correlation between galactic star formation rate and the average SMBH accretion rate in SF galaxies. Recent studies have shown that star formation and AGN accretion are only weakly correlated for individual AGN, but this may be due to the short variability timescale of AGN relative to star formation. Averaging over the full AGN population yields a strong linear correlation between accretion and star formation, consistent with a simple picture in which the growth of SMBHs and their host galaxies are closely linked over galaxy evolution time scales.
\end{abstract}

\firstsection 
\section{Introduction}
For the past decades, it has became increasingly clear that the
formation of stars in galaxies and the growth of supermassive black
holes (SMBHs) might follow parallel evolutionary
paths\citep[see][for a review]{bhreview12}.
A number of studies have directly investigated the link between the
growth of the galactic stellar mass (star formation rate, SFR) and the
growth of SMBH mass (BH accretion rate, BHAR) in galaxies hosting
active galactic nuclei (AGN). For high-luminosity AGNs, an increase in the average SFR as a function
of BHAR has been observed \citep[e.g.][]{lutz08qsosf,serj09qsosf,serj10},
while other studies have also found weak or inverted connections
\citep{page12,harr12}. 
Studies with inclusions of lower luminosity AGNs further suggest that
the evolutionary link between SMBHs and their host galaxies only
exists in high luminosity AGNs that are possibly triggered by mergers,
and there is little or no correlation at lower AGN luminosities \citep[e.g.][]{shao10agnsf,rosa12agnsf,rovi12}.
On the contrary, the studies of the {\it average} BHAR of SF
galaxies imply that the galaxy and SMBH growth rates may be strongly
connected when averaging over the whole population of SF
galaxies \citep{raff11agnsf_aph, mull12agnsf}. Thus, whether BH growth
follows SF in all galaxies, or only in the most powerful systems,
remains a matter of debate. \par
\par

The apparent contradictory results may be attributed to the difference
in the characteristic timescales of SF and BH accretion. 
Both theoretical and observational studies have found evidence
implying that the SMBH accretion rate can vary by several orders of
magnitudes in a time scale much shorter than the typical time scale
for galactic star formation events
\citep[see][and reference therein]{hick13agnvar}.
Therefore, to uncover the relation between the SFR and BHAR, the key
quantity to study may therefore be the {\em average} AGN luminosity of
a population, which thus smoothes over the variations of individual
sources. Here we present a study investigating the average BHAR in SF
galaxies for a sample in the \bootes\ survey region \citep{chen13sfagn,hick13agnvar}. 

\begin{figure}[t]
\begin{center}
 \includegraphics[width=3.0in]{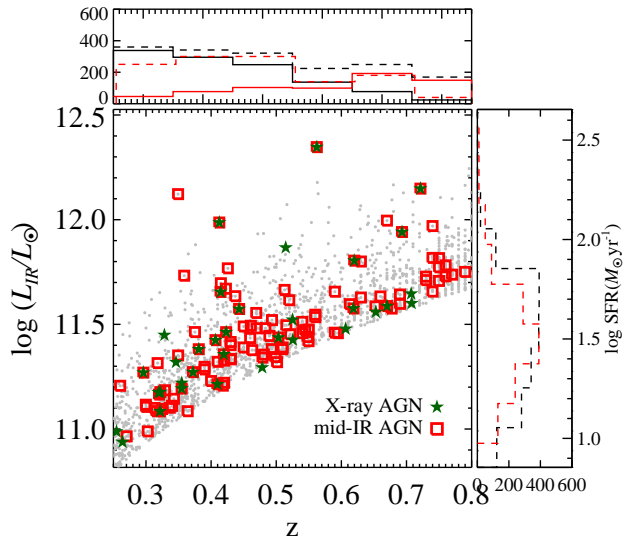} 
\caption{The distribution of redshifts and $\lir\ $ for our sample SF
  galaxies. X-ray AGNs are marked as green stars and mid-IR AGNs are
  marked as red squares. The histograms in redshift and $\lir\ $ are
  also shown in the top and right panels. In the top panel, we show
  the different redshift distributions of the sources with only
  photometric redshifts (solid red line) and the sources with only
  spectroscopic redshifts (solid black line). The redshift
  distributions of AGNs (red dashed line, normalized to scale) and SF
  galaxies are also shown. In the right panel, we show that AGNs (red
  dashed line) and SF galaxies (black dashed line) have similar
  distributions in $\lir\ $ (the histogram of AGNs is normalized to
  scale). These distributions show that the galaxies with identified
  AGN in our sample have distributions in redshift and $\lir\ $
  similar to those of SF galaxies. Figure adapted from \cite{chen13sfagn}.}
\label{fig:lir}
\end{center}
\end{figure}

\section{Sample}
We use data from the 9 deg$^{2}$ \bootes\ multiwavelength survey, which
is covered by the optical observations from NOAO Deep Wide
Field Survey \citep{jann99} and the AGN and
Galaxy Evolution Survey \citep[AGES,][]{AGES11}, X-ray data from the {\it
Chandra} \xbootes Survey \citep{murr05}, mid-IR data from {\it
Spitzer} IRAC Shallow Survey \citep[ISS,][]{eise04},  {\it Spitzer} Deep Wide Field Survey
\citep[SDWFS,][]{ashb09sdwfs} and MIPS $24\;\mu{\rm m}$ observations, and far-IR data from the {\it Herschel}
HerMES survey \citep{hermes}. The redshifts of our sample come from the AGES survey
(spectroscopic) as well as ISS and SDWFS \citep[photometric,][]{brod06photoz}.
Our sample is composed of SF galaxies selected using both the $250\;\mu{\rm m}$ {\it Herschel} SPIRE
observations and the $24\;\mu{\rm m}$ {\it Spitzer} MIPS observations.

\section{Results}
We first derive the SFR for our SF galaxies by extrapolating
the $250\;\mu{\rm m}$ flux measurements to the $8-1000\;\mu{\rm m}$
bolometric infrared luminosity ($\lir\ $) using
the SF galaxy templates from \cite{kirk12}. Since the
bulk of $\lir\ $ comes from the cold dust in SF galaxies,
and the distribution of dust temperature in SF galaxies has
reasonably small scatter, we first derive the total $\lir\ $ for
every galaxy in our sample using the ratio between the observed-frame
monochromatic $250\;\mu{\rm m}$ flux to the total
$\lir\ $ derived from the \cite{kirk12} template. We next convert the
$\lir\ $ to SFR using a Kennicutt relation adjusted to a Chabrier
IMF. The redshift and SFR distribution of our sample is
shown in Fig.~\ref{fig:lir}. We refer the detailed discussion of SFR estimation of our sample to \S2.5 in
\cite{chen13sfagn}. \par

To determine the connection between the SF activity and the {\it
  average} SMBH accretion rate, we measure AGN luminosities using a combination of X-ray and mid-IR
observations. In detail, we convert the mid-IR or
X-ray luminosity for individually detected, X-ray (with $L_X$(2-10keV)
$>10^{42}ergs^{-1}$) or mid-IR selected AGNs \citep{ster05} using the average bolometric
correction factor from \cite{vasu07bolc} and a radiative efficiency of
0.1. For SF galaxies without
direct identified AGN, we measure their average X-ray
luminosity and BHAR using an X-ray stacking analysis in bins of their
SFR \citep{hick07obsagn}. \par
We next measure the average BHAR for our entire sample by averaging
the $L_X$ for the detected AGN and the remaining SF galaxies in bins of
SFR. We find an almost linear relation between SFR and BHAR: 
\begin{equation} \label{eq:avg}
\log (L_{\rm X} [{erg s}^{-1}]) =  
(30.37\pm3.80)+(1.05\pm0.33)\log (L_{IR}/L_{\odot}). 
\end{equation}

\begin{figure}[t]
\begin{center}
 \includegraphics[width=3.4in]{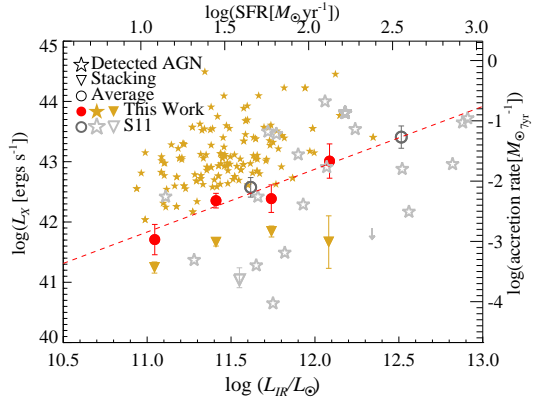} 
\caption{Here we present the SFR-average BHAR relation for far-IR
  selected SF galaxies (circles) from the 9 deg$^2$ \bootes\ field and
  the pencil-beam CDFN \citep{syme11lirlx}. Also plotted are
  individually detected AGNs (stars) and the average BHAR from X-ray
  stacking analysis for normal SF galaxies without direct AGN
  detections. This figure shows that the {\it average} BHAR is
  strongly correlated with the SFR in all rapidly star-forming
  galaxies; and even though the correlation between the $L_{\rm X}$ and
  $\lir\ $ for identified AGNs varies with the depth of the
  observations, the {\it average} correlation is consistent. Figure adapted from \cite{chen13sfagn}.}
\label{fig:avg}
\end{center}
\end{figure}

We tested the effect of the flux limits to the observed
SFR-BHAR correlation using a mock catalog of galaxies with the flux
limits similar to our sample. We found that
for a mock catalog of galaxies without intrinsic SFR-BHAR correlation,
the observed correlation in the flux limits similar to our real
sample is much weaker than the relation observed in
Eq. (1). In addition, the observed evolution of $\lx\ $ as a function of
the average redshift in the different bins of our sample is at least 1.0 dex larger
than the pure redshift evolution of the X-ray luminosity
function, indicating that most, if not all, of our observed trend is due to the intrinsic correlation between SFR and BHAR. \par
We also compared our result with the sample of {\it Herschel} selected
SF galaxies in the pencil-beam {\it Chandra} Deep Field-North
\citep{syme11lirlx} at redshift $z \sim 1$.  Since \cite{syme11lirlx} measured the
$\lir\ $ and $\lx\ $ for both detected AGN and normal SF galaxies
separately, we could make a direct comparison
of the SFR-average BHAR correlation by calculating the average BHAR for
their sample using the same methods that we used to derive our average
BHAR. The results are displayed in Fig.~\ref{fig:avg},
which show the presence of a correlation between SFR and average BHAR
in samples of SF galaxies for a range of redshifts and X-ray flux
limits. 

\section{Conclusion}
In summary, we studied the average BHAR from a sample of SF galaxies
with SFR measurements using {\it Herschel}. We selected AGNs at X-ray
and mid-IR wavelengths to ensure that our BHAR is not biased by AGN
obscuration, and employed an X-ray stacking analysis to measure SMBH
accretion for SF galaxies without direct X-ray or mid-IR
identifications. We obtained an almost linear relation between the
average BHAR and SFR of $\log {\rm BHAR}
=(-3.72\pm0.52)+(1.05\pm0.33)\log{\rm SFR}$, and determined that this
relation also holds for deeper, narrower  observations, suggesting
that the average BHAR to SFR correlation is a universal consequence of
the coevolution between SMBHs and galaxies. The next step of
understanding the SFR to BHAR correlation in different populations of
galaxies requires information on the {\it distribution} of AGN X-ray
luminosity as a function of SFR, which provides motivation for wide,
deep X-ray surveys in multiwavelength fields.


\begin{thebibliography}{}
\bibitem[{Alexander \& Hickox(2012)}]{bhreview12}
Alexander, D.~M., \& Hickox, R.~C. 2012, \textit{New Astronomy Reviews}, 56, 93

\bibitem[{Ashby {et~al.}(2009)Ashby, Stern, Brodwin, Griffith, Eisenhardt,
  Kozłowski, Kochanek, Bock, Borys, Brand, Brown, Cool, Cooray, Croft, Dey,
  Eisenstein, Gonzalez, Gorjian, Grogin, Ivison, Jacob, Jannuzi, Mainzer,
  Moustakas, R\"{o}ttgering, Seymour, Smith, Stanford, Stauffer, Sullivan, van
  Breugel, Willner, \& Wright}]{ashb09sdwfs}
Ashby, M. L.~N., Stern, D., Brodwin, M., {et~al.} 2009, \textit{ApJ}, 701, 428

\bibitem[{Brodwin {et~al.}(2006)Brodwin, Brown, Ashby, Bian, Brand, Dey,
  Eisenhardt, Eisenstein, Gonzalez, Huang, Jannuzi, Kochanek, McKenzie, Pahre,
  Smith, Soifer, Stanford, Stern, Elston, \& Murray}]{brod06photoz}
Brodwin, M., Brown, M. J.~I., Ashby, M. L.~N., {et~al.} 2006, \textit{ApJ}, 651, 791

\bibitem[{Chen {et~al.}(2013)Chen, Hickox, Alberts, Brodwin, Jones, Murray,
  Alexander, Assef, Brown, Dey, Forman, Gorjian, Goulding, {Le Floc'h},
  Jannuzi, Mullaney, \& Pope}]{chen13sfagn}
Chen, C.-T.~J., Hickox, R.~C., Alberts, S., {et~al.} 2013, \textit{ApJ}, 773, 3

\bibitem[{Eisenhardt {et~al.}(2004)Eisenhardt, Stern, Brodwin, Fazio, Rieke,
  Rieke, Werner, Wright, Allen, Arendt, Ashby, Barmby, Forrest, Hora, Huang,
  Huchra, Pahre, Pipher, Reach, Smith, Stauffer, Wang, Willner, Brown, Dey,
  Jannuzi, \& Tiede}]{eise04}
Eisenhardt, P.~R., Stern, D., Brodwin, M., {et~al.} 2004, \textit{ApJS}, 154, 48

\bibitem[{Harrison {et~al.}(2012)Harrison, Alexander, Mullaney, Altieri, Coia,
  Charmandaris, Daddi, Dannerbauer, Dasyra, {Del Moro}, Dickinson, Hickox,
  Ivison, Kartaltepe, {Le Floc'h}, Leiton, Magnelli, Popesso, Rovilos, Rosario,
  \& Swinbank}]{harr12}
Harrison, C.~M., Alexander, D.~M., Mullaney, J.~R., {et~al.} 2012, \textit{ApJ}, 760,
  L15

\bibitem[{Hickox {et~al.}(2007)Hickox, Jones, Forman, Murray, Brodwin, Brown,
  Eisenhardt, Stern, Kochanek, Eisenstein, Cool, Jannuzi, Dey, Brand, Gorjian,
  \& Caldwell}]{hick07obsagn}
Hickox, R.~C., Jones, C., Forman, W.~R., {et~al.} 2007, \textit{ApJ}, 671, 1365

\bibitem[{Hickox {et~al.}(2013)Hickox, Mullaney, Alexander, Chen, Civano,
  Goulding, \& Hainline}]{hick13agnvar}
Hickox, R.~C., Mullaney, J.~R., Alexander, D.~M., {et~al.} 2013, \textit{ApJ}\
  submitted (arXiv:1306.3218)

\bibitem[{Jannuzi \& Dey(1999)}]{jann99}
Jannuzi, B.~T., \& Dey, A. 1999, in ASP Conf.\ Ser.\ 191: Photometric Redshifts
  and the Detection of High Redshift Galaxies, ed. R.~Weymann,
  L.~Storrie-Lombardi, M.~Sawicki, \& R.~Brunner (San Francisco: ASP), 111

\bibitem[{Kirkpatrick {et~al.}(2012)Kirkpatrick, Pope, Alexander, Charmandaris,
  Daddi, Dickinson, Elbaz, Gabor, Hwang, Ivison, Mullaney, Pannella, Scott,
  Altieri, Aussel, Bournaud, Buat, Coia, Dannerbauer, Dasyra, Kartaltepe,
  Leiton, Lin, Magdis, Magnelli, Morrison, Popesso, \& Valtchanov}]{kirk12}
Kirkpatrick, A., Pope, A., Alexander, D.~M., {et~al.} 2012, \textit{ApJ}, 759, 139

\bibitem[{Kochanek {et~al.}(2012)Kochanek, Eisenstein, Cool, Caldwell, Assef,
  Jannuzi, Jones, Murray, Forman, Dey, Brown, Eisenhardt, Gonzalez, Green, \&
  Stern}]{AGES11}
Kochanek, C.~S., Eisenstein, D.~J., Cool, R.~J., {et~al.} 2012, \textit{ApJ}s, 200, 8

\bibitem[{{Lutz} {et~al.}(2008){Lutz}, {Sturm}, {Tacconi}, {Valiante},
  {Schweitzer}, {Netzer}, {Maiolino}, {Andreani}, {Shemmer}, \&
  {Veilleux}}]{lutz08qsosf}
{Lutz}, D., {Sturm}, E., {Tacconi}, L.~J., {et~al.} 2008, \textit{ApJ}, 684, 853

\bibitem[{Mullaney {et~al.}(2012{\natexlab{a}})Mullaney, Pannella, Daddi,
  Alexander, Elbaz, Hickox, Bournaud, Altieri, Aussel, Coia, Dannerbauer,
  Dasyra, Dickinson, Hwang, Kartaltepe, Leiton, Magdis, Magnelli, Popesso,
  Valtchanov, Bauer, Brandt, {Del Moro}, Hanish, Ivison, Juneau, Luo, Lutz,
  Sargent, Scott, \& Xue}]{mull12agnsf}
Mullaney, J.~R., Pannella, M., Daddi, E., {et~al.} 2012{\natexlab{a}}, \textit{MNRAS},
  419, 95

\bibitem[{Murray {et~al.}(2005)Murray, Kenter, Forman, Jones, Green, Kochanek,
  Vikhlinin, Fabricant, Fazio, Brand, Brown, Dey, Jannuzi, Najita, McNamara,
  Shields, \& Rieke}]{murr05}
Murray, S.~S., Kenter, A., Forman, W.~R., {et~al.} 2005, \textit{ApJ}s, 161, 1

\bibitem[{Oliver {et~al.}(2012)Oliver, Bock, Altieri, Amblard, Arumugam,
  Aussel, Babbedge, Beelen, B\'{e}thermin, Blain, Boselli, Bridge, Brisbin,
  Buat, Burgarella, Castro-Rodr\'{\i}guez, Cava, Chanial, Cirasuolo, Clements,
  Conley, Conversi, Cooray, Dowell, Dubois, Dwek, Dye, Eales, Elbaz, Farrah,
  Feltre, Ferrero, Fiolet, Fox, Franceschini, Gear, Giovannoli, Glenn, Gong,
  Solares, Griffin, Halpern, Harwit, Hatziminaoglou, Heinis, Hurley, Hwang,
  Hyde, Ibar, Ilbert, Isaak, Ivison, Lagache, Floc'h, Levenson, Faro, Lu,
  Madden, Maffei, Magdis, Mainetti, Marchetti, Marsden, Marshall, Mortier,
  Nguyen, O'Halloran, Omont, Page, Panuzzo, Papageorgiou, Patel, Pearson,
  P\'{e}rez-Fournon, Pohlen, Rawlings, Raymond, Rigopoulou, Riguccini, Rizzo,
  Rodighiero, Roseboom, Rowan-Robinson, Portal, Schulz, Scott, Seymour, Shupe,
  Smith, Stevens, Symeonidis, Trichas, Tugwell, Vaccari, Valtchanov, Vieira,
  Viero, Vigroux, Wang, Ward, Wardlow, Wright, Xu, \& Zemcov}]{hermes}
Oliver, S.~J., Bock, J., Altieri, B., {et~al.} 2012, \textit{MNRAS}, 23

\bibitem[{Page {et~al.}(2012)Page, Symeonidis, Vieira, Altieri, Amblard,
  Arumugam, Aussel, Babbedge, Blain, Bock, Boselli, Buat,
  Castro-Rodr\'{\i}guez, Cava, Chanial, Clements, Conley, Conversi, Cooray,
  Dowell, Dubois, Dunlop, Dwek, Dye, Eales, Elbaz, Farrah, Fox, Franceschini,
  Gear, Glenn, Griffin, Halpern, Hatziminaoglou, Ibar, Isaak, Ivison, Lagache,
  Levenson, Lu, Madden, Maffei, Mainetti, Marchetti, Nguyen, O'Halloran,
  Oliver, Omont, Panuzzo, Papageorgiou, Pearson, P\'{e}rez-Fournon, Pohlen,
  Rawlings, Rigopoulou, Riguccini, Rizzo, Rodighiero, Roseboom, Rowan-Robinson,
  {S\'{a}nchez Portal}, Schulz, Scott, Seymour, Shupe, Smith, Stevens, Trichas,
  Tugwell, Vaccari, Valtchanov, Viero, Vigroux, Wang, Ward, Wright, Xu, \&
  Zemcov}]{page12}
Page, M.~J., Symeonidis, M., Vieira, J.~D., {et~al.} 2012, \textit{Nature}, 485, 213

\bibitem[{Rafferty {et~al.}(2011)Rafferty, Brandt, Alexander, Xue, Bauer,
  Lehmer, Luo, \& Papovich}]{raff11agnsf_aph}
Rafferty, D.~A., Brandt, W.~N., Alexander, D.~M., {et~al.} 2011, \textit{ApJ}, 742, 3

\bibitem[{Rosario {et~al.}(2012)Rosario, Santini, Lutz, Shao, Maiolino,
  Alexander, Altieri, Andreani, Aussel, Bauer, Berta, Bongiovanni, Brandt,
  Brusa, Cepa, Cimatti, Cox, Daddi, Elbaz, Fontana, {F\"{o}rster Schreiber},
  Genzel, Grazian, {Le Floch}, Magnelli, Mainieri, Netzer, Nordon, {P\'{e}rez
  Garcia}, Poglitsch, Popesso, Pozzi, Riguccini, Rodighiero, Salvato,
  Sanchez-Portal, Sturm, Tacconi, Valtchanov, \& Wuyts}]{rosa12agnsf}
Rosario, D., Santini, P., Lutz, D., {et~al.} 2012, \textit{A\&A}, 545, A45

\bibitem[{Rovilos {et~al.}(2012)Rovilos, Comastri, Gilli, Georgantopoulos,
  Ranalli, Vignali, Lusso, Cappelluti, Zamorani, Elbaz, Dickinson, Hwang,
  Charmandaris, Ivison, Merloni, Daddi, Carrera, Brandt, Mullaney, Scott,
  Alexander, {Del Moro}, Morrison, Murphy, Altieri, Aussel, Dannerbauer,
  Kartaltepe, Leiton, Magdis, Magnelli, Popesso, \& Valtchanov}]{rovi12}
Rovilos, E., Comastri, A., Gilli, R., {et~al.} 2012, \textit{A\&A}, 546, A58

\bibitem[{Serjeant \& Hatziminaoglou(2009)}]{serj09qsosf}
Serjeant, S., \& Hatziminaoglou, E. 2009, \textit{MNRAS}, 397, 265

\bibitem[{Serjeant {et~al.}(2010)Serjeant, Bertoldi, Blain, Clements, Cooray,
  Danese, Dunlop, Dunne, Eales, Falder, Hatziminaoglou, Hughes, Ibar, Jarvis,
  Lawrence, Lee, Michałowski, Negrello, Omont, Page, Pearson, van~der Werf,
  White, Amblard, Auld, Baes, Bonfield, Burgarella, Buttiglione, Cava, Dariush,
  de~Zotti, Dye, Frayer, Fritz, Gonzalez-Nuevo, Herranz, Ivison, Lagache,
  Leeuw, Lopez-Caniego, Maddox, Pascale, Pohlen, Rigby, Rodighiero, Samui,
  Sibthorpe, Smith, Temi, Thompson, Valtchanov, \& Verma}]{serj10}
Serjeant, S., Bertoldi, F., Blain, A.~W., {et~al.} 2010, \textit{A\&A}, 518, L7

\bibitem[{Shao {et~al.}(2010)Shao, Lutz, Nordon, Maiolino, Alexander, Altieri,
  Andreani, Aussel, Bauer, Berta, Bongiovanni, Brandt, Brusa, Cava, Cepa,
  Cimatti, Daddi, Dominguez-Sanchez, Elbaz, {F\"{o}rster Schreiber}, Geis,
  Genzel, Grazian, Gruppioni, Magdis, Magnelli, Mainieri, {P\'{e}rez
  Garc\'{\i}a}, Poglitsch, Popesso, Pozzi, Riguccini, Rodighiero, Rovilos,
  Saintonge, Salvato, {Sanchez Portal}, Santini, Sturm, Tacconi, Valtchanov,
  Wetzstein, \& Wieprecht}]{shao10agnsf}
Shao, L., Lutz, D., Nordon, R., {et~al.} 2010, \textit{A\&A}, 518, L26+

\bibitem[{Stern {et~al.}(2005)Stern, Eisenhardt, Gorjian, Kochanek, Caldwell,
  Eisenstein, Brodwin, Brown, Cool, Dey, Green, Jannuzi, Murray, Pahre, \&
  Willner}]{ster05}
Stern, D., Eisenhardt, P., Gorjian, V., {et~al.} 2005, \textit{ApJ}, 631, 163

\bibitem[{Symeonidis {et~al.}(2011)Symeonidis, Georgakakis, Seymour, Auld,
  Bock, Brisbin, Buat, Burgarella, Chanial, Clements, Cooray, Eales, Farrah,
  Franceschini, Glenn, Griffin, Hatziminaoglou, Ibar, Ivison, Mortier, Oliver,
  Page, Papageorgiou, Pearson, P\'{e}rez-Fournon, Pohlen, Rawlings, Raymond,
  Rodighiero, Roseboom, Rowan-Robinson, Scott, Smith, Tugwell, Vaccari, Vieira,
  Vigroux, Wang, \& Wright}]{syme11lirlx}
Symeonidis, M., Georgakakis, A., Seymour, N., {et~al.} 2011, \textit{MNRAS}, 417, 2239

\bibitem[{Vasudevan \& Fabian(2007)}]{vasu07bolc}
Vasudevan, R.~V., \& Fabian, A.~C. 2007, \textit{MNRAS}, 381, 1235
\end{thebibliography}
\end{document}